\documentclass[twocolumn,prd,showpacs,showkeys,
amsmath,amssymb]{revtex4}
\usepackage{graphicx}
\renewcommand{\vec}[1]{\bm{#1}}
\begin{document}

\title{Limitations on the attainable intensity of high power lasers}
\author{A.M. Fedotov}
\author{N.B. Narozhny}%
\affiliation{National Research Nuclear University MEPhI, Moscow}
\author{G. Mourou}
\affiliation{ Laboratoire d'Optique Appliqu\'{e}e, Ecole
Polytechnique, Paris}
\author{G. Korn}
\affiliation{Max Plank Institute for Quantum Optics, Garching}

\begin{abstract}
It is shown that even a single $e^-e^+$ pair created by a super
strong laser field in vacuum would cause development of an
avalanche-like QED cascade which rapidly depletes the incoming laser
pulse. This confirms the old N. Bohr conjecture that the electric
field of the critical QED strength $E_S=m^2c^3/e\hbar$ could never
be created.
\end{abstract}

\maketitle

F. Sauter was the first who had introduced the critical QED field
$E_S=m^2c^3/e\hbar =1.32\cdot 10^{16}\mbox{V/cm}$ in his remarkable
papers \cite{Sauter} where he considered the so-called Klein paradox
\cite{Klein}. The content of the paradox is that, according to the
Dirac electron theory, an electron can penetrate into a potential
barrier which is greater than twice the rest energy of the electron.
Sauter was also the first who had interpreted the Klein paradox as
creation of $e^-e^+$ pairs by an external field from vacuum.

Even before Sauter, N. Bohr has suggested a postulate that the
electric field of the strength $E_S$ could never be created in
principle, see Chapter IV $\S$ 11 in A. Sommerfeld book \cite{Somm}.
However, further development of QED has demonstrated that no
difficulties arised in the theory at $E\sim E_S$. Klein paradox was
successively resolved \cite{Sauter,NN,Nik}, the nonlinear correction
to electromagnetic Lagrangian valid at arbitrary field strength was
calculated \cite{HE,Schw}, many authors theoretically considered
physics at field strength higher than the critical value. Thus the
Bohr's conjecture was completely forgotten.

It was shown recently \cite{Nar2004,Nar2006,Fed} that pairs can be
created in vacuum by focused laser pulses with the peak field
strength below the critical value. Simultaneous focusing of multiple
colliding pulses may lower the threshold field strength for the
process down to $10^{-2}E_S$ \cite{few_pulses}. The corresponding
optical laser intensity is of the order
$5\cdot10^{25}\mbox{W/cm}^2$. Projects (e.g., ELI, XFEL) to achieve
such intensity in the near future are under way already
\cite{Ring,TM,ELI}, so that pair creation from vacuum by an external
electromagnetic fields, thought as a sort of gedanken experiment for
years, may become accessible for experimental observation very soon.

Dependence of the average number of created pairs on field strength
is determined by the exponential factor $\exp(-\pi E_S/E)$ and is
very sharp. At $E\sim E_S$ the estimated number of produced pairs
$N_{e^-e^+}$ is so huge that their total rest energy becomes
comparable with the energy of the laser pulse itself
\cite{Nar2004,Nar2006,Fed}. Certainly, $N_{e^-e^+}$ at $E\sim E_S$
was overestimated in Refs.~\cite{Nar2004,Nar2006,Fed} because it was
obtained under assumption that the laser field could be considered
as a given classical background. However, that result indicates that
the process of pair creation leads to depletion of the laser field
and the effect of back-reaction should be taken into account.
Moreover, this estimate could serve an argument in favor of Bohr's
postulate on unattainability of pair creating electromagnetic field
with $E=E_S$.

But this point is not the end of the story. There exists another,
and even more effective, mechanism for depletion of a pair creating
laser pulse. The point is that the created electron and positron can
be very quickly accelerated by the laser field to relativistic
energies and emit hard photons, which produce then new $e^-e^+$
pairs. These effects have been already observed in the famous E144
SLAC experiment \cite{SLAC2}, but yet just as single events, because
the energy of electrons and hard photons, as well as the laser
intensity were not high enough. At high laser intensities
interaction of the created electron and positron with the laser
field can lead to production of multiple new particles and thus to
formation of an avalanche-like electromagnetic cascade
\cite{BellKirk,GC,BellKirk1}. Such cascade have many features in
common with the cascades produced as the result of a high-energy
particle interacting with dense matter. The latter were well studied
both experimentally and theoretically, see, e.g.,
Ref.~\cite{recent}. However, there exists an important distinctive
feature of the laser-induced cascades, as compared with the air
showers arising due to primary cosmic ray entering atmosphere. In
our case the laser field plays not only the role of a target
(similar to a nuclei in the case of air showers). It is responsible
also for acceleration of slow particles \cite{BellKirk}, playing
thus the role of a linac in the SLAC experiment. Thus, the
laser-induced cascade in vacuum looks very much like electron
avalanche which can occur due to impact ionization in
dielectric-filled trench used for electrical isolation of
semiconductor devices \cite{sem}. In this letter we will consider
the mechanism of onset and development of an electromagnetic cascade
initiated by a pair created in vacuum by short focused laser pulses.
We will show that creation of even a single pair may result in
complete destruction of the laser field.

We will consider an ultra-relativistic ($a_0=eE/m\omega c\gg1$) but
still subcritical ($E,H\ll E_S$) laser field. The formation length
for quantum processes in such a field $l_f\sim\lambda/a_0$
\cite{Ritus} satisfies the condition $l_f\ll\lambda$, where
$\lambda$ is both the wavelength and the characteristic length of
variation of the laser field. Then the total probabilities of the
processes of photon emission by a charged particle and pair creation
by a photon depend on local values of three dimensionless
parameters: two field invariants $\mathcal{F},\mathcal{G}$ and the
dynamical parameter $\chi$,
$$\mathcal{F}=\frac{E^2-H^2}{2E_S^2},\,\,\mathcal{G}=\frac{\vec{E}
\cdot\vec{H}}{E_S^2},\,\,
\chi=\frac{e\hbar}{m^3c^4}\sqrt{-(F_{\mu\nu}p^\nu)^2},$$ where
$p^{\nu}$ is 4-momentum either of electron (positron), or photon.
The parameter $\chi$ for an electron is exactly its proper
acceleration in the field measured in Compton units $mc^3/\hbar$. In
the lab frame it can be expressed by
$$\chi_e=\frac{\sqrt{f_\parallel^2+\gamma_e^2 f_\perp^2}}{eE_S},$$
where $\vec{f}_\parallel$ and $\vec{f}_\perp$ are the components of
the Lorentz force, longitudinal and transverse to the electron
motion respectively, $\gamma_e$ is the electron Lorentz factor.

If $\mathcal{F},\mathcal{G}\ll\chi$, and $a_0\gg1$ any laser field
can be considered locally as a constant crossed field \cite{Ritus}.
Indeed, the probability of any process $W$ can be approximated by
$W(\mathcal{F},\mathcal{G},\chi)\approx W(0,0,\chi)$, which is the
probability of the same process in a plane wave field. Since the
condition $l_f\ll\lambda$ is satisfied, $W$ is the probability rate
of the process in a constant crossed field. The rates for both
photon emission and pair creation by photon in such field can be
easily found in literature, see, e.g., \cite{Ritus}. The total
probability rates in the limiting case of large $\chi$ read
\begin{equation}\label{W_lim}
W_{e,\gamma}\sim\frac{\alpha m^2c^4}{\hbar\epsilon_{e,\gamma}}
\chi_{e,\gamma}^{2/3},
\end{equation}
where $\epsilon_e$, $\epsilon_\gamma$ denote the energy of the
initial particle. In the classical limit $\chi_\gamma\ll 1$ the rate
of pair creation by a photon is exponentially suppressed by the
factor $\exp(-8/3\chi_\gamma)$.

Particles of the created pair can be considered to be at rest
initially $\chi_e(0)=E/E_S\ll1$. The cascade can develop if the
particles are able to emit hard photons with $\chi_\gamma\gtrsim 1$,
i.e. if $\chi_e\gtrsim 1$. This means that the act of hard photon
emission can take place if the field accelerates particles and the
time of acceleration $t_{acc}$ is small enough, $t_{acc}\ll t_{esc}$
at least, where $t_{esc}$ is the time of stay of the particle in the
laser pulse. For a laser pulse focused up to the diffraction limit
the time $t_{esc}$ may be estimated to be $t_{esc}\sim\lambda/2c$.

\begin{figure}
\includegraphics[width=8cm]{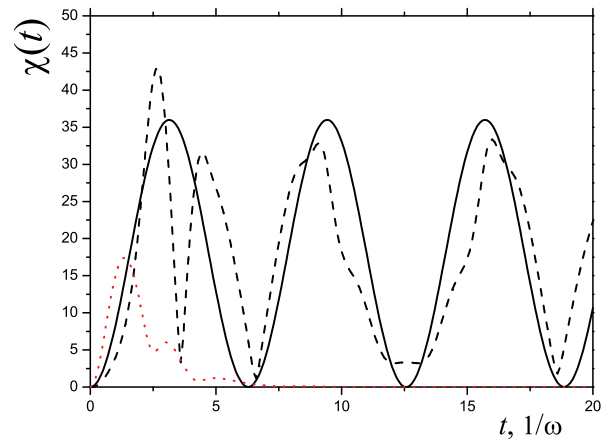}
\caption{\label{fig:chi(t)} Evolution of quantum dynamical parameter
$\chi$ along the particle trajectory for $a_0=3\cdot 10^3$, $\hbar\omega=
1\mbox{eV}$ in three cases:
head-on collision of two elliptically polarized plane waves (solid line);
collision at $90^0$ of two linearly polarized plane waves with orthogonal
linear polarizations (dash line); single tightly focused e-polarized laser
beam (dot line).}
\end{figure}

To estimate $t_{acc}$, we will use the model of a uniform purely
electric field rotating with frequency $\omega$. Such field can be
realized practically in the anti-nodes of a circularly polarized
monochromatic standing wave. The initial electron and positron will
be accelerated by the field in opposite directions. We will consider
below only one branch of the cascade initiated by the positron.
Since no hard photons can be emitted at $t<t_{acc}$, we will treat
the motion of the particle classically. The equation of motion
$\dot{\vec{p}}_e(t)=e\vec{E}(t)$ can be easily solved. The result is
presented by solid line in Fig.~\ref{fig:chi(t)}. Parameter $\chi_e$
in this case is not conserved as, e.g., in a constant field but
oscillates with period $2\pi/\omega$ and the amplitude of
oscillations $\chi_{e\,max}\approx 2(E/E_S)a_0$ can considerably
exceed unity even for relatively moderate field intensities.
Fig.~\ref{fig:chi(t)} contains also the results of numerical
calculations for the cases of crossed orthogonally polarized plane
waves and $e$-polarized tightly focused laser beam. It is clear that
although the details of evolution of $\chi_e(t)$ are much more
complicated in general case, the basic qualitative features, such as
the period and the amplitude of oscillations are of the same order
as in the case of a uniformly rotating electric field.

Consider parameter $\chi_e(t)$ at $t\ll \pi/\omega$. It is
determined by the product of the positron Lorentz factor, in our
case $\gamma_e\sim eEt/mc$, and the transverse component of the
field. It can be easily observed that the angle between the field
and particle momentum increases as $\theta_e=\omega t/2$.  Thus, we
estimate $E_\perp\sim E\omega t$ and come to
\begin{equation}\label{chi_appr}
\chi_e(t)\sim\left(\frac{E}{E_S}\right)^2 \frac{mc^2\omega t^2}{\hbar}.
\end{equation}
The crucial point here is that the rotating field not only
accelerates the particle but also contorts its trajectory, so that
the Lorentz force gains transverse component. As it is seen from the
Eq.~(\ref{chi_appr}), $t_{acc}$ can be estimated as
\begin{equation}\label{t_acc}
t_{acc}\sim \frac{\hbar}{\alpha mc^2\mu}\sqrt{\frac{mc^2}{\hbar\omega}},
\end{equation}
where $\mu=E/E_*$ and $E_*=\alpha E_S\approx E_S/137$. For optical
frequency, this time period remains just a small fraction of the
rotation period provided that  $I> 10^{24}\mbox{W/cm$^2$}$.

The positron radiation lifetime (mean free path/c) $t_e$ with
respect to photon emission can be estimated by $t_e\sim W_e^{-1}$.
Thus, at the moment of photon emission we have
\begin{equation}\label{t_eq_te}
\epsilon_e\sim eE\frac{c}{W_e},\quad \chi_e\sim
\left(\frac{E}{E_S}\right)^2 \frac{mc^2\omega }{\hbar W_e^2}.
\end{equation}
After substitution of (\ref{t_eq_te}) into Eq.~(\ref{W_lim}), we
find that $\chi_e\sim\mu^{3/2}$ and
\begin{equation}\label{main}
\epsilon_{e}\sim mc^2\mu^{3/4}\sqrt{\frac{mc^2}{\hbar\omega}},\quad
t_e\sim\frac{\hbar}{\alpha mc^2\mu^{1/4}}\sqrt{\frac{mc^2}{\hbar\omega}}.
\end{equation}
Since the energy of a photon emitted by an ultrarelativistic
positron with $\chi_e\gtrsim$ is of the order
$\epsilon_\gamma\sim\epsilon_e$, $\chi_\gamma\sim\chi_e$, the photon
lifetime with respect to pair production $t_\gamma$ is of the order
$t_e$.

It can be easily seen from Eqs.~(\ref{t_acc}) and (\ref{main}) that
the following hierarchy of time scales
\begin{equation}\label{hierarchy}
t_{acc}\lesssim t_e,t_{\gamma}\ll t_{esc}
\end{equation}
is respected if
\begin{equation}\label{conds}
\mu\gtrsim 1, \quad \mu^{1/4}\gg\frac{1}{\alpha}
\sqrt{\frac{\hbar\omega}{mc^2}}\,.
\end{equation}

Eq.~(\ref{hierarchy}) determines the necessary conditions for
occurrence of electromagnetic cascade. Indeed, if $t_e\gtrsim
t_{acc}$, then $\chi_\gamma$ of the emitted photon can be $\gtrsim
1$ and hence it may create a pair. On the other hand, the number of
successive events of photon emission and pair production throughout
the time period $t_{esc}$ is large if $t_e,t_\gamma\ll t_{esc}$.

It is worth noting that for optical frequencies the second
restriction in Eq.~(\ref{conds}) is weaker than the first one.
Therefore the necessary conditions for occurrence of the cascade is
reduced to the relation $\mu\gtrsim 1$ in this case. Consequently
the field strength $E_*$ determines a natural threshold for
electromagnetic cascades. Such field performs the work $\sim mc^2$
over an electron at its radiation free path. The corresponding laser
intensity is $I_*\approx 2.5\cdot 10^{25}\mbox{W/cm$^2$}$. Hence,
the cascade will occur at intensities of the order of
$10^{25}\mbox{W/cm$^2$}$ or higher.

\begin{figure}[t]
\includegraphics[width=8.5cm]{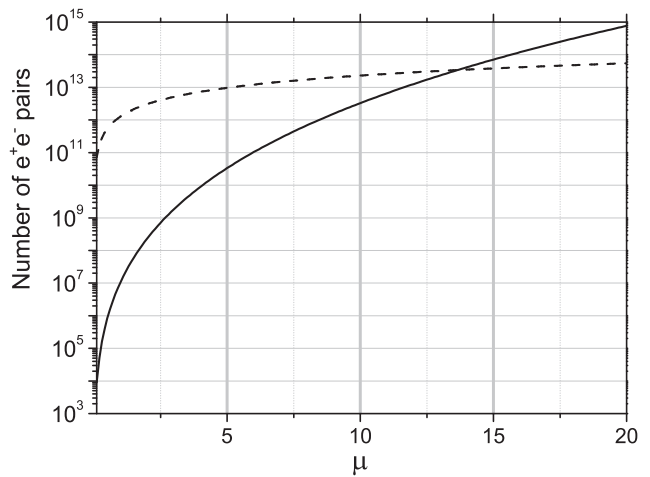}
\caption{\label{fig:Nee} The number of created pairs $N_e$ (solid curve)
versus parameter $\mu$. The dash line shows the limit for
$N_e$ determined by the energy of the laser pulse. The laser
frequency $\hbar\omega=1eV$.}
\end{figure}

The total number of pairs created per one laser shot can be estimated as
\begin{equation}\label{Ncasc}
 N_e\sim\exp\left(\frac{t_{esc}}{t_e}\right)\sim \exp\left[\pi\alpha
 \mu^{1/4}\sqrt{\frac{mc^2}{\hbar\omega}}\right],
\end{equation}
see the solid curve in Fig.~\ref{fig:Nee}.

The most important result of our estimation is that the number of
created pairs is so huge that their net energy can become even equal
to the energy stored in the laser pulse. Under assumption that the
laser pulse is focused to diffraction limit, so that the volume of
the focal area where pairs are created is $\sim(\lambda/2)^3$, the
total energy of laser field can be estimated by $W\sim (E^2/4\pi)
(\lambda/2)^3$. Thus, the maximum number of pairs that can be
created is restricted by the value
\begin{equation}\label{Nmax}
N_{e,max}=\frac{W}{2\epsilon_\epsilon}\sim
\alpha\mu^{5/4} \left(\frac{mc^2}{\hbar\omega}\right)^{5/2}.
\end{equation}
This quantity is represented by a dash line in Fig.~\ref{fig:Nee}.
We see that the net energy of created pairs becomes of the order of
the energy of the laser pulse already at $\mu\approx 10$. Such value
of $\mu$ will be attained in collision of two circularly polarized
laser pulses of intensity $I\approx 6\cdot 10^{26}\mbox{W/cm}^2$.

\begin{figure}[t]
\includegraphics[width=8.5cm]{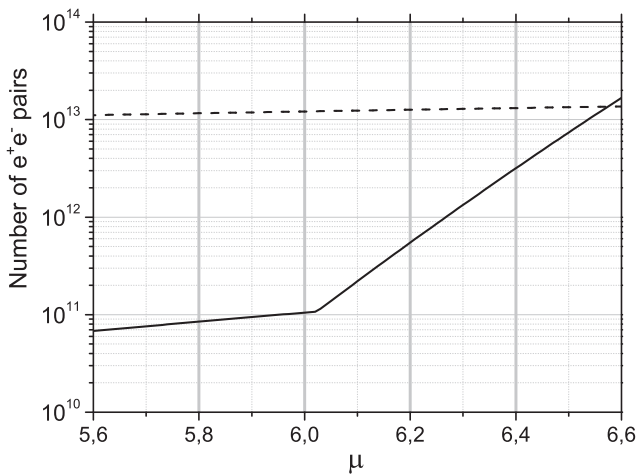}
\caption{\label{fig:Nee1} The number of created pairs $N_e$ (solid curve)
versus parameter $\mu$ with due regard for the number of pairs created
by two colliding circularly polarized $10$fs laser pulses with
$\hbar\omega=1\rm eV$. The point of the curve break corresponds to the
threshold value of $\mu$ when only one pair is created. The dash line
shows the limit for $N_e$ determined by the energy of the laser pulse.}
\end{figure}

This estimation was obtained under assumption that only one pair was
created per one shot. However, according to Ref.~\cite{Nar2006} the
threshold intensity for pair creation by two colliding circularly
polarized $10$fs laser pulses with $\hbar\omega=1\rm eV$ is equal to
$I_{th}\approx 2.3\cdot 10^{26}\mbox{W/cm}^2$. The dependence of the
number of created pairs $N_e$ on the intensity of colliding pulses
is very sharp, and at $I\approx 6\cdot 10^{26}\mbox{W/cm}^2$ it
reaches the value of $N_e\approx 6\cdot 10^8$. This means that
destruction of the laser pulse will take place much earlier then it
follows from Fig.~\ref{fig:Nee}. Dependence of the total number of
created pairs with due regard for the number of pairs initially
created by the laser field is presented in Fig.~\ref{fig:Nee1}. The
break of the curve takes place at the threshold value of
$\mu=\mu_{th}$ which is equal approximately to $\mu_{th}\approx6$
for the case of two colliding circularly polarized laser pulses
according to Ref.~\cite{Nar2006}. It is assumed that for
$\mu<\mu_{th}$ a particle initiating the cascade was not created due
to the pair creation process but was situated there initially. The
latter assumption is true for Fig.~\ref{fig:Nee} as well.

We see that $N_e$ reaches the maximum possible value at $\mu\approx
6.6$ which corresponds to intensity of colliding pulses $I\approx
2.7\cdot 10^{26}\mbox{W/cm}^2$.

To conclude, we have described a new kind of QED cascades that can
be initiated in the laser focus by even a single pair created by the
laser field from vacuum. The multiplicity of such cascade is no
longer limited by the initial energy of a particle initiating it.
Rather, it is limited by the time of stay of particles in laser
field, or, under more extreme conditions, by the total
electromagnetic energy stored in the pulse. In the latter case,
development of the cascade causes depletion of the laser pulse, thus
imposing strict limitation on the maximum attainable laser
intensity.

In this letter, we have based our consideration on a model of
rotating electric field. To obtain more realistic results one needs
solve the system of 3D kinetic equations taking into account
back-reaction of the cascade development on the laser field. At the
moment, we are working on this program. However, our present
estimations convincingly confirm Bohr's $80$-years old conjecture
that the critical QED field $E_S$ can be never attained. Note that
our conclusion applies to the electric field strength in the
reference frame, where the magnetic field either vanishes or
$\vec{H}\parallel\vec{E}$.

We are grateful to V. Mur, V. Popov, S. Kelner, E. Echkina, D. Rogozkin, D. Voskresensky, I. Kostyukov, E. Nerush, J. Rafelski, S.V. Bulanov, A. Bell, J.
Kirk, I. Sokolov and G. R\"{o}pke for fruitful discussions. A.~M.~F.
is especially grateful to Prof. H.~Ruhl for hospitality at LMU
Munich and stimulating discussions on implementation of QED cascades
in numerical simulations. This work was partially supported by the
grants RFBR~09-02-01201-a, RFBR~09-02-12201-ofi$\_$m and
RNP~2.1.1/1871.

\end{document}